\documentclass[amsmath,amssymb,aps,pra,superscriptaddress,10pt]{revtex4-1}

\usepackage{amsmath}
\usepackage{bm}
\usepackage{geometry}
\usepackage[dvipdfmx]{graphicx}
\usepackage[caption=false]{subfig}
\usepackage{epstopdf}
\usepackage{refcount}
\usepackage{array}
\usepackage{hyperref}
\usepackage[dvipsnames]{xcolor}
\begin{document}
\title{Splitting instability of a doubly quantized vortex in superfluid Fermi gases}

\author{W. Van Alphen}
\affiliation{TQC, Universiteit Antwerpen, Universiteitsplein 1, B-2610 Antwerpen, Belgium}
\author{H. Takeuchi}
\affiliation{Department of Physics and Nambu Yoichiro Institute of Theoretical and Experimental Physics (NITEP), \color{black}Osaka Metropolitan University\color{black}, Osaka 558-8585, Japan}
\author{J. Tempere}
\affiliation{TQC, Universiteit Antwerpen, Universiteitsplein 1, B-2610 Antwerpen, Belgium}
\affiliation{Lyman Laboratory of Physics, Harvard University, Cambridge, Massachusetts 02138, USA}

\begin{abstract}
The splitting instability of a doubly-quantized vortex in the BEC-BCS
crossover of a superfluid Fermi gas is investigated by means of a low-energy effective field theory. Our linear stability analysis and
non-equilibrium numerical simulations reveal that the character of the
instability drastically changes across the crossover. In the BEC-limit,
the splitting of the vortex into two singly-quantized vortices occurs
through the emission of phonons, while such an emission is
completely absent in the BCS-limit. In the crossover-regime, the
instability and phonon emission are enhanced, and the lifetime of a doubly-quantized vortex
becomes minimal. The emitted phonon
 \color{black}
 can be observed as a spiraling pattern amplified due to the rotational superradiance, known as a mechanism to carry away energy and angular momentum from a spinning black hole.
 \color{black}
We also
investigate the influence of temperature, population imbalance, and
three-dimensional effects.
\end{abstract}
\maketitle
\section*{Introduction}
\label{sec:intro}
An understanding of the dynamics of quantized vortices is essential to understand the behavior of superfluids \cite{THDonnelly,THAnnett,THPethick,THFetter} such as superfluid helium, superconductors, quantum gases or nucleonic superfluids.
Vortices with two or more circulation quanta are known to be energetically unstable with respect to splitting into singly-quantized vortices \cite{THPethick}.
Vortex decay via splitting is a nontrivial process which has thus far been observed dynamically only in superfluid quantum gases \cite{EXPShin} thanks to the high level of control and tunability of these systems.
Theoretically, the splitting of doubly quantized vortices (DQVs) in Bose-Einstein condensates (BECs) at zero temperature has mainly been investigated \color{black} by solving the Bogoliubov equations \color{black}
\cite{THPu,THSkryabin,THSimula,THMottonen,THLund2,THKawaguchi,THHuh,THLund,THFukuyama,THNilsen,THHiroSplit}.
 While this splitting instability exhibits a complicated finite-size effect by coupling to collective excitations \cite{THPu,THMottonen,THKawaguchi,THHuh,THLund, THHiroSplit}, its experimental evidence in uniform superfluids is still lacking, partly because the instability is quite weak in uniform systems \citep{THweaksplit,THHiroSplit,THAranson}.
 Superfluid Fermi gases have a much richer phenomenology of elementary and collective excitations than their bosonic counterparts, and this should be reflected in the vortex decay dynamics.
\color{black}
Multiply quantized vortices have been also studied in superconductors \cite{PhysRevLett.81.2783,PhysRevLett.85.1528} and fermionic superfluids in the weak-pairing BCS regime \cite{PhysRevLett.119.067003}.
\color{black}
Nevertheless, vortex decay in superfluid Fermi gases remains largely unexplored, mainly due to the fact that hydrodynamic models for these Fermi superfluids are still under development \cite{THRanderiaSaDeMelo,THSimonucciStrinati,THYeong,THManiniSalasnich}.

In this paper, we study 
the splitting instability of a DQV in the entire BEC-BCS crossover of a superfluid Fermi gas based on a recently developed low-energy effective field theory (EFT) \cite{THKTLDEpjB,THKTVPrA94, THKliminNJP,THKTVPrA94, THKliminNJP}. 
The lifetime of the DQV and the dynamics of the instability are investigated for a uniform, cylindrically trapped Fermi superfluid \cite{EXPMukherjee}.
We show that the instability is enhanced in the crossover regime, and can be observed experimentally through a spiraling phonon pattern amplified due to
 \color{black}
 the rotational superradiance known to occur in spinning black holes, detected very recently \cite{Cui2023}.
 \color{black}
 Finally, we also analyze the effects of temperature and population imbalance on the instability. 


\section*{Theoretical model}
\label{sec:model}

The system under consideration is an ultracold Fermi gas in which particles of mass $m$ and opposite pseudo-spin interact via a contact potential with $s$-wave scattering length $a_s$. In the context of the EFT, this system can be described in terms of a superfluid order parameter $\Psi(\mathbf{r},t)$, representing the bosonic field of Cooper pairs. Under the assumption that this field varies slowly around the bulk value in both space and time, a gradient expansion of the Euclidean-time action functional of the fermionic system can be performed, resulting in 
the following three-dimensional (3D) equation of motion:
\begin{equation}
i \tilde{D}(\vert \Psi \vert^2) \frac{\partial \Psi}{\partial t} = -C \, \nabla_{\mathbf{r}}^2 \Psi + Q \frac{\partial^2 \Psi}{\partial t^2} + \left( \mathcal{A}(\vert \Psi \vert^2) + 2  E \, \nabla_{\mathbf{r}}^2 \vert \Psi \vert^2 - 2  R \frac{\partial^2 \vert \Psi \vert^2}{\partial t^2} \right) \Psi. \label{eq:eqofmot}
\end{equation}
This equation is a type of non-linear Schr\"{o}dinger equation which is closely related to both the Gross-Pitaevskii equation for Bose-Einstein condensates \cite{THKTVPrA94} and the Ginzburg-Landau equation for BCS superfluids \cite{THRanderiaSaDeMelo}.  We use the natural units of $\hbar = 1$, $2m = 1$, $E_F = 1$. A detailed overview of the model can be found in Ref.~\cite{THKTLDEpjB} or in the supplemental material \cite{SM},
 together with the analytical expressions for
 $\mathcal{A}$, $C$, $\tilde{D}$, $E$, $Q$ and $R$ in terms of the inverse temperature $\beta$, the average chemical potential $\mu$, the imbalance chemical potential $\zeta$
 \footnote{
 $\mu $ and $\zeta $ are defined in terms of the chemical
 potentials of the spin-up and spin-down populations as
  \color{black} $\mu = (\mu_{\uparrow} + \mu_{\downarrow})/2$ \color{black} and
  \color{black} $\zeta = (\mu_{\uparrow} - \mu_{\downarrow})/2$ \color{black}.
 }
 ,
\color{black}
 and the bulk amplitude $\vert \Psi_{\infty} \vert$ (i.e. the superfluid gap $\Delta$).
 All our results depend on $(k_F a_s)^{-1}$ (with $k_F$ the Fermi wave number) only through $\mu/\Delta$.
 The relation between $\mu/\Delta$ and $(k_F a_s)^{-1}$ changes depending on which equation of state (EOS) is chosen.
 Here we choose the EOS based on the mean-field approximation
\footnote{\color{black}
The results by using another EOS (e.g., quantum Monte Carlo \cite{PhysRevLett.93.200404,PhysRevA.85.051601} or the experiment \cite{nascimbene2010exploring}) would be reproduced after rescaling of $(k_F a_s)^{-1}$ \cite{PhysRevA.100.063634}.
\color{black}
}.
\color{black}
 The coefficients $\tilde{D}$ and $\mathcal{A}$ depend fully upon the local amplitude $\vert \Psi(\mathbf{r},t) \vert$ \cite{THKTDPrA}.
We assign to $\vert \Psi_{\infty} \vert$ and $\mu$ the mean-field values that are obtained by simultaneously solving the saddle-point gap and number equations \cite{THDevreeseTempere}.

The stationary solution for a doubly-quantized vortex can be represented in polar coordinates $(r,\phi,z)$ as
\begin{equation}
\label{eq:statsol}
\Psi_s(r,\phi) = f(r) e^{i l \phi},
\end{equation}
where the amplitude $f(r)$ only depends on the radial coordinate, and $l=2$.
 \color{black}
 Such a vortex state is feasible as was demonstrated convincingly by manipulating vortices in Fermi superfluids in box-shaped, toroidal traps \cite{Pace2022}.
 \color{black}
 By substituting \eqref{eq:statsol} into \eqref{eq:eqofmot}, one can find a numerical solution for $f(r)$. It is convenient to express the length scale in units of the healing length $\xi$, which is a measure for the width of the vortex. An analytic expression for $\xi$ can be derived through a variational ansatz for the stationary vortex solution and a minimization of the EFT free energy \cite{SM,THBookChapter}. For a typical experimental setup $k_F \sim 0.5$ $\mu$m, this yields $\xi \approx$ 1 $\mu$m, 800 nm, 10 $\mu$m for $(k_F a_s)^{-1} = 2, 0, -2$ respectively.

The main assumption of the EFT model is that the order parameter $\Psi(\mathbf{r},t)$ varies slowly in both space and time \cite{THKTLDEpjB}. This corresponds to the conditions that the pair field should vary over a spatial region larger than the pair correlation length, and that the energy of the fluctuations remains below the pair-breaking threshold ($2 \Delta$ in the BCS-regime, $2 \sqrt{\Delta^2 + \mu^2}$ in the BEC-regime). A detailed study of the validity of the model reveals that the theory is
\color{black}
less reliable for describing dark solitons in the BCS-regime at low temperatures \cite{THLvAKTPrA}, where $\Delta$ becomes small and the ratio of the pair correlation length $\xi_{\rm pair}$ to the healing length is close to unity.
This is also the case with singly quantized vortex ($l=1$).
On the other hand, in our case of a doubly quantized vortex ($l=2$), we have typically the condition $\xi>\xi_{\rm pair}$ since the healing length for $l=2$ is twice or more than that for $l=1$ \cite{SM}.
\color{black}

The dynamic stability of a DQV in a Fermi superfluid can be studied by adding a small complex perturbation to the stationary vortex solution:
\begin{equation}
\Psi(\mathbf{r},t) = \Big( f(r) + \Phi(\mathbf{r},t) \Big) e^{i l \phi}.
\label{eq:factorphase}
\end{equation}
Small excitations of the system can be described by a fluctuation field of the form \cite{THLund2}
\begin{equation}
\label{eq:planewave}
\Phi(\textbf{r},t) = \phi_1(r) e^{i(m\phi +k_z z - \omega t)} + \phi_2^*(r) e^{-i(m\phi + k_z z  - \omega^* t)},
\end{equation}
where $m$ is an angular momentum quantum number and $k_z$ is the wave number along the $z$-axis.
The equation of motion \eqref{eq:eqofmot} can then be linearized with respect to the perturbation amplitudes $\phi_1$ and $\phi_2$, which leads to differential equations of the following form:
\begin{align}
\label{eq:DQuv1}
\alpha_1(r) \frac{\partial^2 \phi_1}{\partial r^2} + &\alpha_2(r) \frac{\partial \phi_1}{\partial r} + \Big( \omega^2 \, \alpha_3(r) + \omega \, \alpha_4(r) + \alpha_{5,+}(r) \Big) \phi_1 \notag \\
&+ \alpha_6(r) \frac{\partial^2 \phi_2}{\partial r^2} + \alpha_7(r) \frac{\partial \phi_2}{\partial r} + \Big( \omega^2 \, \alpha_8(r) + \alpha_9(r) \Big) \phi_2 = 0,
\end{align}
\begin{align}
\label{eq:DQuv2}
\alpha_1(r) \frac{\partial^2 \phi_2}{\partial r^2} + &\alpha_2(r) \frac{\partial \phi_2}{\partial r} + \Big( \omega^2 \, \alpha_3(r) - \omega \, \alpha_4(r) + \alpha_{5,-}(r) \Big) \phi_2 \notag \\*
&+ \alpha_6(r) \frac{\partial^2 \phi_1}{\partial r^2} + \alpha_7(r) \frac{\partial \phi_1}{\partial r} + \Big( \omega^2 \, \alpha_8(r) + \alpha_9(r) \Big) \phi_1 = 0.
\end{align}
The expressions for the position-dependent coefficients $\alpha_i(r)$ are given in the supplemental material \cite{SM}. Because of the centrifugal term ($\propto (l \pm m)^2/r^2$) in the expressions for $\alpha_{5,\pm}$, $\phi_1$ and $\phi_2$ are only allowed to be finite at the core center ($r=0$) for $m=-l$ and $m=l$, respectively. In all other cases, $\phi_1$ and $\phi_2$ must vanish at the center. As is the case for a DQV in BECs, the splitting instability is induced by the so-called core mode, a collective mode which is localized around the vortex core \cite{THLund}. We thus restrict our analysis to the case of $m=\pm 2$.


\section*{Linear stability analysis}

\begin{figure}[htbp]
\centering
\centerline{
\includegraphics[scale=0.35]{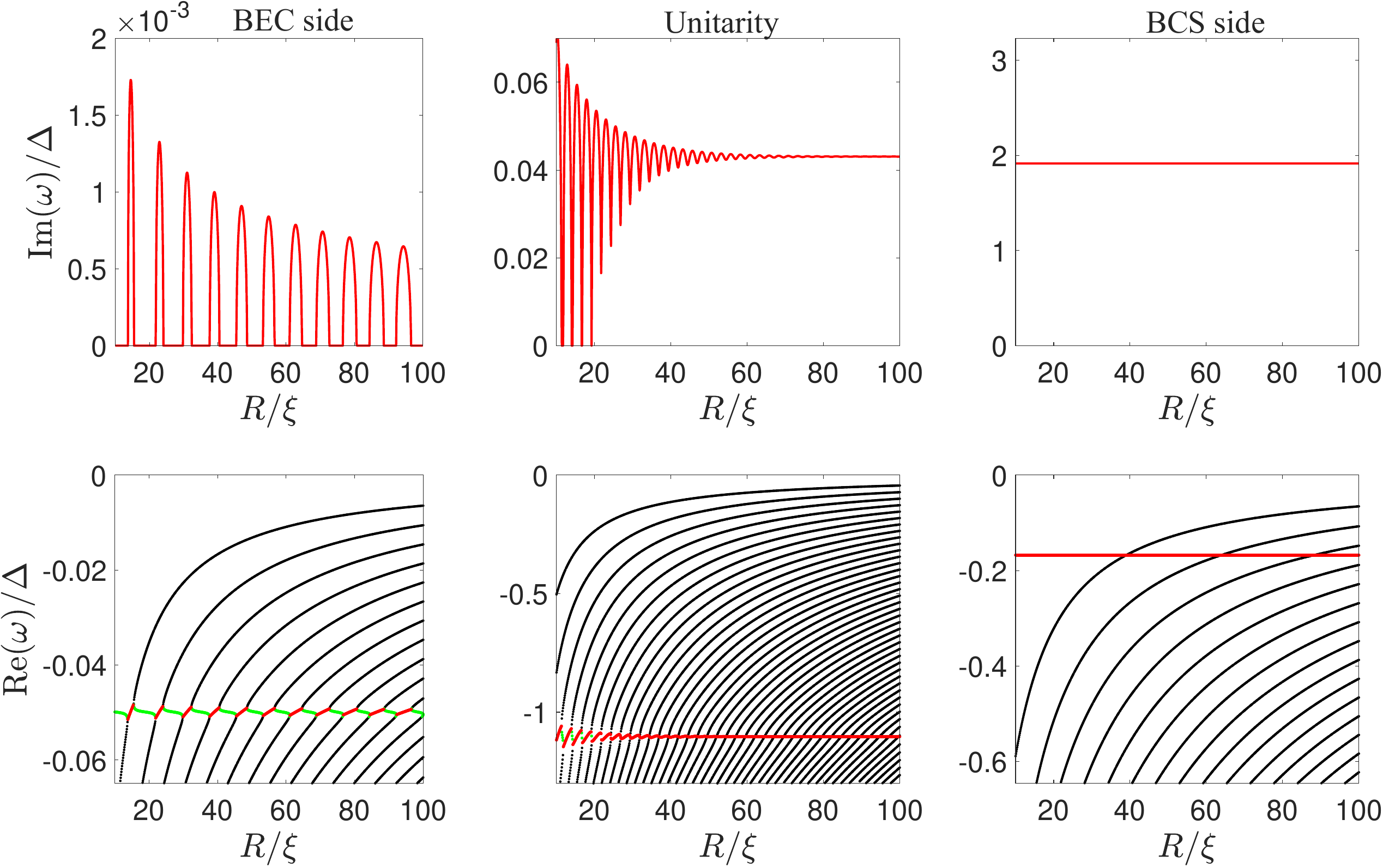}}
\caption{Imaginary (upper row) and real (lower row) part of the eigenfrequencies $\omega/\Delta$ of the excitation modes of a doubly quantized vortex in function of the system size $R/\xi$, for $(k_F a_s)^{-1} = 2$ (left column), $(k_F a_s)^{-1} = 0$ (middle column) and $(k_F a_s)^{-1} = -2$ (right column). Eigenvalues corresponding to non-localized modes are colored black, eigenvalues corresponding to stable core modes are colored green, and eigenvalues corresponding to unstable core modes are colored red.}
\label{fig:CoreRadius}
\end{figure}

We will first study the splitting instability \color{black} at $T\approx 0$ ($\beta=10^3 \times E_F^{-1}$) \color{black} without imbalance \color{black} by assuming that the fluid and possible excitations are homogeneous in the $z$ direction, \color{black} which comes down to setting $k_z = 0$ and $\zeta = 0$
 \footnote{\color{black}
 The three dimensional deformation of the vortex line can be important when the system size
  along the z axis is larger than $2\pi/k_{z,c} \approx 20\xi,~8\xi,~4\xi$ for $(k_F a_s)^{-1}=-2,~0,~2$, respectively \cite{SM}.
 \color{black}}.
  The upper left panel of Figure \ref{fig:CoreRadius} shows the numerical result for the imaginary part of the complex eigenvalues $\text{Im}(\omega)/\Delta$ in function of the radial system size $R$, for $(k_F a_s)^{-1} = 2$ (BEC-regime).
\color{black}
Here, we imposed the Neumann boundary condition at $r=R$.
\color{black}
 Whenever $\text{Im}(\omega) \neq 0$, the DQV is unstable with respect to splitting into two SQVs. The DQV's lifetime is then related to the inverse of $\text{Im}(\omega)$. The graph demonstrates that the instability only occurs within certain intervals of values for $R$. 
This periodic change in the stability of the vortex in function of $R$ has also been predicted in the case of bosonic superfluids \cite{THHiroSplit}, and can be explained by considering the real part $\text{Re}(\omega)$ in the lower left panel of Figure \ref{fig:CoreRadius}. 
Eigenvalues corresponding to radially propagating phonon modes are colored black, while eigenvalues corresponding to core modes are colored green for stable core modes (no imaginary part) and red for unstable core modes. One can observe that, in order for the DQV to decay, the core mode must come into resonance with a phonon mode that can carry away energy and angular momentum from the vortex.
Since the radial momentum of the phonons is quantized in the finite system, the DQV's lifetime is expected to strongly depend on the system size. As $R$ increases, the spectrum of phonon levels becomes more dense, reducing the regions of stability, until eventually, in the limit $R \rightarrow \infty$, the phonon spectrum will become
a continuum and the imaginary eigenvalue is expected to take on a constant (nonzero) value \cite{THHiroSplit}.

The middle column of Figure \ref{fig:CoreRadius} shows the eigenvalues for $(k_F a_s)^{-1} = 0$ (unitarity). One can observe in the lower panel that the ratio of the core mode energy to the gap has increased with respect to the BEC-regime. As a consequence, the core mode encounters a much denser spectrum of phonon modes to couple with, and the oscillations
of the imaginary part of the complex eigenvalue in function of $R$  quickly disappear. Hence, at unitarity, the finite-size effect of the vortex instability vanishes for much smaller system sizes than in the BEC-limit.

Finally, the right column of Figure \ref{fig:CoreRadius} shows the imaginary and real parts of the eigenmodes in function of $R$ for $(k_F a_s)^{-1} = -2$ (BCS-regime). In contrast to the BEC- and crossover regime, the core mode is observed to be permanently unstable with a constant non-zero imaginary part, indicating that the lifetime of the DQV is insensitive to the system size on the BCS-side. The fact that the core mode doesn't seem to interact with the phonon modes at all implies that some other kind of mechanism induces the instability here. Analytically,  we find that, in the deep BCS-regime, where the coefficients $Q$ and $R$ become large and the coefficient $\tilde{D}$ can be neglected \cite{VATaTePRA}, the linear equations \eqref{eq:DQuv1} and \eqref{eq:DQuv2} can be reduced to a Schr\"{o}dinger-like equation with eigenvalue $\omega^2$. 
The core mode then plays the role of a bound state of the potential created by the vortex profile, and the instability is induced solely by the core mode with $\omega^2 < 0$.

A possible microscopic explanation beyond the EFT for the behavior of the instability in the BCS-regime is that, instead of the collective excitations, the core mode now couples to the single-particle excitations of the system (i.e.\ unpaired fermions), which play a more significant role on this side of the interaction domain. The presence of these unpaired particles is taken into account 
through the local value of the single-particle excitation spectrum $E_{\mathbf{q}}(r) = \sqrt{f^2(r) + (\mathbf{q}^2 - \mu)}$ in the EFT coefficients $\tilde{D}_s$, $\mathcal{A}$ and $\partial_s \mathcal{A}_s$ (where $\mathbf{q}$ represents the wave vector of the fermionic modes) . Close to the vortex core, the amplitude $f(r)$ of the stationary vortex solution goes to zero, meaning $E_{\mathbf{q}}(r)$ will decrease as well. Consequently, the core mode, which is exactly localized around this region, might be able to couple to the single-particle excitation modes to induce the decay \color{black} through, e.g., the pair-breaking process. This sort of coupling with single-particle excitations is important to understand the dissipative mechanism in the inelastic collisions of dark solitons \cite{scott2012decay}, which is qualitatively described in our EFT description \cite{THvALKTColl}. \color{black}

\section*{Population imbalance and finite temperatures}

\begin{figure}[htbp]
\centering
\centerline{
\includegraphics[scale=0.8]{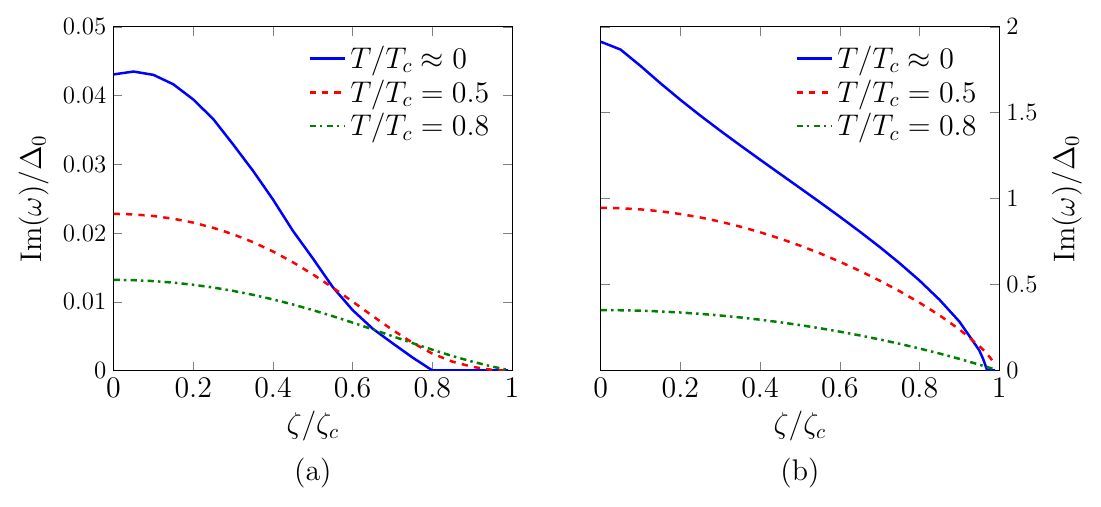}}
\caption{Imaginary part of the complex eigenfrequency of a doubly quantized vortex in function of the imbalance chemical potential $\zeta/\zeta_c$, for several values of the temperature $T/T_c$, for $R = 250\xi$, and for (a) $(k_F a_s)^{-1} = 0$ and (b) $(k_F a_s)^{-1} = -2$.}
\label{fig:DQimomZetaBeta}
\end{figure}

By tuning the parameters $\beta$ and $\zeta$, the EFT analysis allows to investigate the effects of temperature and imbalance on the unstable mode and the DQV's lifetime. Since both of these parameters tend to have only small effects on the BEC-side of the interaction domain \cite{THvALKTColl,THLvAKTSI}, we focus on their impact in the crossover- and BCS-regime. Figure \ref{fig:DQimomZetaBeta} shows the imaginary part $\text{Im}[\omega]/\Delta_0$ in function of the imbalance parameter $\zeta/\zeta_c$, for several values of $T/T_c$ and for several values of the interaction parameter. Here, $\Delta_0$ is the superfluid gap for $T = \zeta = 0$, while $\zeta_c$ and $T_c$ indicate the critical values of the imbalance parameter and the temperature for the phase transition to the normal state, respectively. It is clear that increasing the population imbalance \color{black} typically \color{black} makes the value $\text{Im}(\omega)$ decrease, which in turn means that the lifetime of the DQV will increase. Hence, imbalanced fermionic superfluids could allow us to control the splitting stability. Increasing the temperature of the system appears to have the same result, except very close to the critical value of the imbalance parameter
 \footnote{
 \color{black}
The current work considers the impact of temperature only on the dynamic instability of a DQV. The dissipative dynamics due to thermal excitations is an important subject for the future work. Some Gross-Pitaevskii models at finite temperatures include such a dissipative effect by replacing the real coefficient of the first-order differentiation of time by a complex one, corresponds to $\tilde{D}$ in our model. The coefficient $\tilde{D}$ is real, but instead the EFT includes the effect of dissipation through all the coefficients in Eq. (1).
\color{black}
 }.
In earlier work, similar effects have been observed for the dynamic instability of dark solitons in superfluid Fermi gases \cite{THLvAKTSI}. In that case, it was argued that the stabilization is due to the unpaired particles that fill up the core of the solitary excitation as the imbalance or temperature are increased.
The same kind of reasoning can be applied to the vortex core.

\section*{Non-equilibrium dynamics}

To study the full non-equilibrium dynamics of the splitting instability beyond the linear regime, we used the EFT's non-linear equation of motion \eqref{eq:eqofmot} to perform numerical simulations of the zero-temperature decay of the DQV
\color{black}
 in a Fermi superfluid in a cylindrical trap with a hard wall at $r=R$.
\color{black}
 The time evolution is carried out by discretizing the space-time grid and applying a finite-difference fourth order Runge-Kutta (RK4) algorithm \cite{SM,VATaTePRA}. For the present calculations, the spatial and temporal resolution are taken to be respectively $5\%$ of $\xi$ and $2\%$ of $t_F = \omega_F^{-1} = (E_F/\hbar)^{-1}$.  A small amount of random noise with a fixed amplitude \color{black} ($0.01\times \vert \Psi_\infty \vert $) \color{black} is added to the initial vortex state in order to trigger the instability. The DQV's lifetime can then be defined as the moment at which two separate SQV cores can be resolved at a distance $\xi$ from each other, similar to how it was characterized in the experiment in Ref.\ \cite{EXPShin}.

\begin{figure}[htbp]
\centering
\centerline{
\includegraphics[scale=0.35]{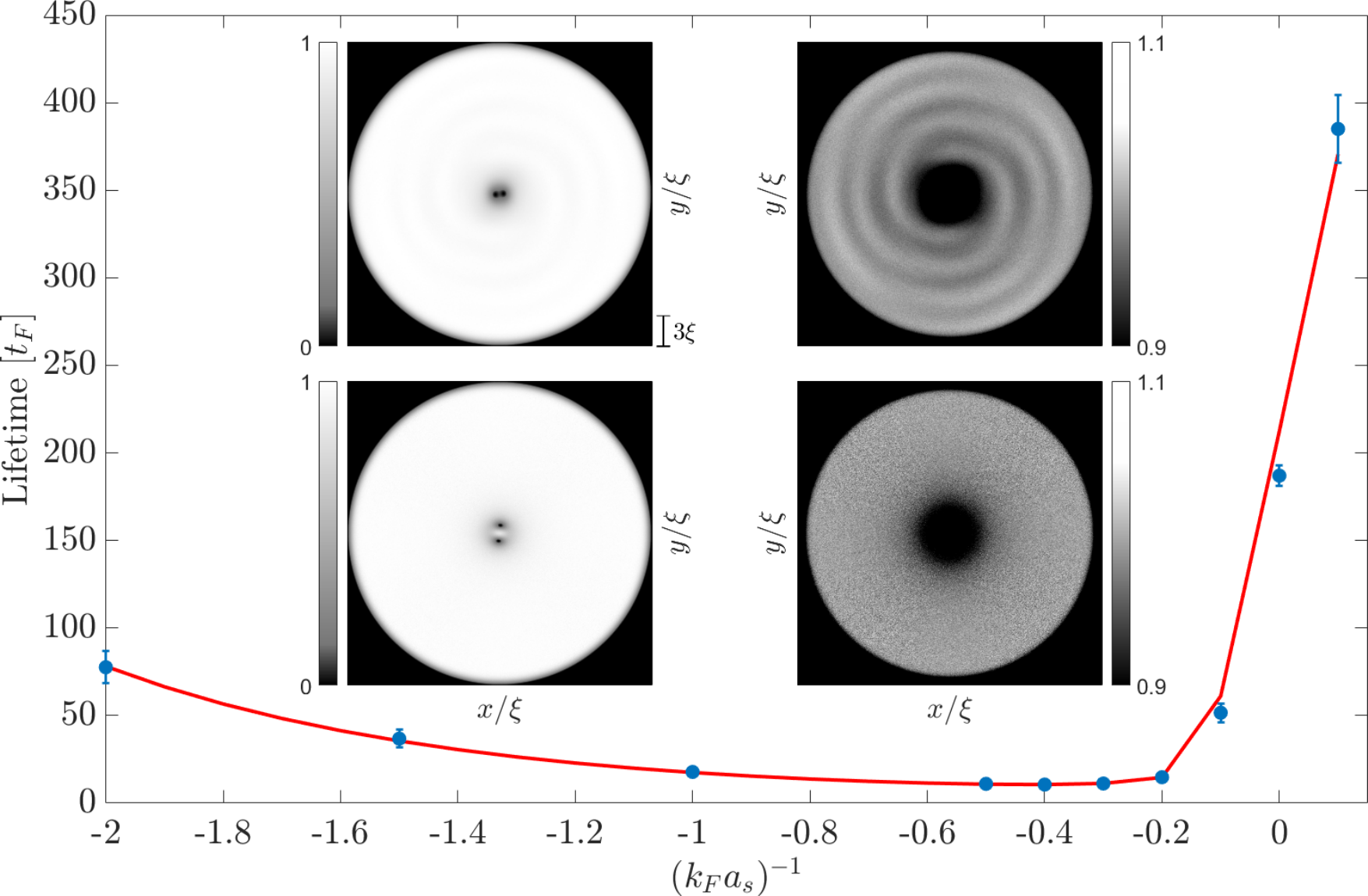}}
\caption{Comparison of the numerical result for the lifetime of a DQV in a cylindrical box (blue dots) to the scaled values $1/\text{Im}(\omega)$ predicted by the linear stability analysis (red line). Every result of the numerical simulations is obtained as the average of five runs, with a standard deviation depicted by the error bars. The insets shows snapshots of the relative pair density $\vert \Psi \vert^2/\vert \Psi_{\infty} \vert^2$ during the decay for $(k_F a_s)^{-1} =0$ at $t = 230 \, t_F$  (upper row) and $(k_F a_s)^{-1} =-2$ at $t = 140 \, t_F$ (lower row).}
\label{fig:LifeTime}
\end{figure}

The blue dots in Figure \ref{fig:LifeTime} show the result for the lifetime of the DQV 
as a function of $(k_F a_s)^{-1}$. The lifetime of the vortex starts to increase very steeply towards the BEC-side, making it more difficult to detect and study the dynamics of the splitting instability in the deep BEC-limit \cite{THAranson}. In the crossover regime, on the other hand, the instability seems to be strongly enhanced as the lifetime reaches its minimal value. The results of the numerical simulations can also be compared to the predictions of the linear stability analysis, estimating the lifetime as $\propto 1/\text{Im}(\omega)$ (since the vortex-vortex distance grows exponentially as \color{black} $\propto e^{\text{Im}(\omega) t}$\color{black}). After scaling with a constant factor $A \approx 6.94$, the graphs of $A/\text{Im}(\omega)$ and the lifetime are found to be in very good agreement.

The insets of Figure \ref{fig:LifeTime} show snapshots of the pair field density during the DQV's decay in the crossover regime (upper images) and BCS-regime (lower images). The left images show the pair density between zero and the bulk value $\vert \Psi_{\infty} \vert$, while the right images only show values of the density in a close range around the bulk value, in order to make the phonons in the system more apparent.
In the crossover regime, one can clearly observe that the splitting of the DQV is accompanied by the emission of spiraling phonons. This is in accordance with the predictions of the linear analysis that, in the BEC- and crossover-regime, the instability is induced by a coupling of the core mode to radially propagating phonon modes with positive angular momentum \footnote{See also Fig.\ 3a of Ref.\ \cite{THHiroSplit} for a schematic of the phonon emission in BEC systems.}. 
In the BCS-regime, on the other hand, no phonons are found to be emitted during the decay process, which again agrees with the earlier result that the instability is induced solely by the core mode for $(k_F a_s)^{-1} = -2$.

The spiral pattern is amplified over time by rotational superradiance \cite{SuperRad1}, as was discussed \color{black} in the context of the splitting instability \cite{THHiroSplit,SuperRad2,Patrick2022}. \color{black} The superradiance is a possible mechanism to extract energy from a spinning body or black hole by spontaneous emission and amplification of electromagnetic waves (see also \cite{SuperRad,SuperRad3}),
\color{black} similarly to the Penrose process \cite{penrose1971extraction} \color{black}.
According to \color{black}Unruh's \color{black} theory of the acoustic metric \cite{SuperRad4}, the superradiance of phonons can happen in superfluids \cite{Calogeracos1999,Volovik,Slatyer2005,PhysRevA.73.033604,Takeuchi2008}.
In our system, the phonon is emitted spontaneously \color{black} and amplified \color{black} in the splitting instability \color{black} by extracting the energy and angular momentum outward\color{black}.
\color{black} While the \color{black} superradiance has been observed in a classical system \cite{SuperRad6}, our system is an appealing candidate to simulate the black hole physics in quantum systems.

\section*{Conclusions}

In this work, the splitting instability of a DQV in a uniform superfluid Fermi gas was investigated by means of a low-energy effective field theory. Our linear stability analysis revealed that, on the BEC-side of the crossover, a DQV is unstable against splitting into two SQVs when the core mode of the vortex couples to phonon modes. As a result, the vortex lifetime depends strongly on the size of the system. In the BCS-regime, on the other hand, the lifetime becomes insensitive to this finite-size effect. Full numerical simulations of the decay of a DQV in a uniformly trapped Fermi superfluid confirmed these predictions, and demonstrated that the lifetime is minimal in the crossover regime.  
A study of the effect of temperature and population imbalance on the splitting instability revealed that tuning the values of these parameters allows one to adjust the strength of the instability, providing experimentalists with more control over the timing and course of the decay process.
The lifetime of a DQV at unitarity,  $\sim $ 10 ms for a typical experimental setup $k_F^{-1} \sim$ 0.5 $\mu$m, is short enough to observe the splitting instability and the rotational superradiance as a spiraling phonon. Such experimental observation will be valuable for developing the non-equilibrium dynamics of fermionic superfluids and simulating black hole physics in a quantum system.

\acknowledgments W.\ Van Alphen acknowledges financial support in the form of a Ph.\ D.\ fellowship of the Research Foundation - Flanders (FWO). This research was supported by the University Research Fund (BOF) of the University of Antwerp and by the Flemish Research Foundation (FWO-Vl),
\color{black}
projects G.0429.15.N, GOH11.22N, G.0618.20.N, G.0608.20.N.
\color{black}
 H. Takeuchi was supported by JSPS KAKENHI Grant Numbers JP17K05549,
JP18KK0391, JP20H01842), and in part by the OCU "Think globally, act
locally" Research Grant for Young Scientists 2019 and 2020 through the
hometown donation fund of Osaka City.




\newpage
\section*{Supplemental material}
\def\thesection{Appendix \Alph{section}}
\renewcommand{\thesection}{S\arabic{section}}
\setcounter{section}{0}
\renewcommand{\thesubsection}{S\arabic{subsection}}
\setcounter{subsection}{0}
\renewcommand{\theequation}{S\arabic{equation}}
\setcounter{equation}{0}
\renewcommand{\thefigure}{S\arabic{figure}}
\setcounter{figure}{0}

 \subsection*{Overview of the EFT}
 \label{sec:appEFT}
 In this section we provide a brief overview of the EFT model and the expressions for the EFT expansion coefficients. More detailed derivations and explanations can be found in Ref.\ \cite{THDevreeseTempere, THKTLDEpjB, THLombardiPhD}.

 The system of interest is an ultracold, dilute Fermi gas, in which particles of opposite pseudo-spin interact via an $s$-wave contact potential. The Euclidian-time action functional of this system can be written down in terms of the fermionic (Grassmann) fields $\psi_{\sigma}(\mathbf{x},\tau)$ and $\bar{\psi}_{\sigma}(\mathbf{x},\tau)$:
 \begin{multline}
 S[\psi] = \int_0^\beta d\tau \int d\textbf{x} \left[\sum_{\sigma\in\lbrace \uparrow,\downarrow\rbrace}\bar{\psi}_\sigma(\textbf{x},\tau)\left(\frac{\partial}{\partial \tau} -\nabla^2_{\textbf{x}}-\mu_\sigma\right)\psi_\sigma(\textbf{x},\tau)  +g \, \bar{\psi}_\uparrow(\textbf{x},\tau)\bar{\psi}_\downarrow(\textbf{x},\tau)\psi_\downarrow(\textbf{x},\tau)\psi_\uparrow(\textbf{x},\tau)\right]
 \label{eq:FermionicAction}
 \end{multline}
 where $g$ is the strength of the contact interaction and the label $\sigma$ denotes the spin degree of freedom. The quartic interaction term can be decoupled through the Hubbard-Stratonovich (HS) transformation, which introduces the bosonic pair field $\Psi(\mathbf{x},\tau)$ (the HS field is often also denoted as $\Delta$, but here we use $\Psi$ to emphasize its interpretation as a position- and time-dependent order parameter for the system) \cite{THDevreeseTempere}. The fermionic degrees of freedom can then be integrated out. If we assume that the pair field $\Psi(\mathbf{x},\tau)$ only varies slowly around its constant background value $\Psi_{\infty}$, we can perform a gradient expansion around $\Psi_{\infty}$ up to second order in the spatial and temporal derivatives of $\Psi(\mathbf{x},\tau)$ \cite{THKTLDEpjB}. This results in the following Euclidian-time effective action functional for the bosonic pair field:
 \begin{align}
 S_{\text{EFT}}[\Psi ]= \int_{0}^{\beta} d\tau \int d\mathbf{r} &\left[ \frac{D}{2}\left( \bar{\Psi}\frac{\partial \Psi}{\partial \tau}- \frac{\partial \bar{\Psi}}{\partial \tau }\Psi \right) +  \Omega_s + C  \left( \nabla_{\mathbf{r}} \bar{\Psi} \cdot\nabla_{\mathbf{r}} \Psi \right) - E \left( \nabla_{\mathbf{r}} \vert \Psi \vert^2 \right)^2 \right. \notag \\
 &+ \left. Q \frac{\partial \bar{\Psi}}{\partial \tau} \frac{\partial \Psi}{\partial \tau} - R \left( \frac{\partial \vert \Psi \vert^2}{\partial \tau} \right)^2 \right] \label{eq:action}
 \end{align}
 This effective action functional forms the starting point for our study of the snake instability in the main work. The thermodynamic potential $\Omega_s$ is given by:
 \begin{align}
 \label{eq:omega}
 \Omega_s &= -\frac{1}{8 \pi k_F a_s}\vert \Psi \vert^2 - \int \frac{d\mathbf{k}}{(2 \pi)^3} \left\lbrace \frac{1}{\beta} \ln[2 \cosh(\beta E_{\mathbf{k}}) + 2 \cosh(\beta \zeta)] - \xi_{\mathbf{k}} - \frac{\vert \Psi \vert^2}{2 k^2}  \right\rbrace
 \end{align}
 while the gradient expansion coefficients $D$, $C$, $E$, $Q$ and $R$ are defined as
 \begin{align}
 D &= \int \frac{d \mathbf{k}}{(2 \pi)^3} \frac{\xi_{\mathbf{k}}}{\vert \Psi \vert^2} [ f_1(\beta,\xi_{\mathbf{k}},\zeta) - f_1(\beta,E_{\mathbf{k}},\zeta) ]  \label{eq:d} \\
 C &= \int \frac{d \mathbf{k}}{(2 \pi)^3} \frac{k^2}{3 m} f_2 (\beta,E_{\mathbf{k}},\zeta) \label{eq:c} \\
 E &= 2 \int \frac{d \mathbf{k}}{(2 \pi)^3} \frac{k^2}{3 m} \, \xi_{\mathbf{k}}^2 \, f_4(\beta,E_{\mathbf{k}},\zeta)  \label{eq:e} \\
 Q &= \frac{1}{2 \vert \Psi \vert^2} \int \frac{d \mathbf{k}}{(2 \pi)^3} [f_1(\beta,E_{\mathbf{k}},\zeta) -(E_{\mathbf{k}}^2 + \xi_{\mathbf{k}}^2) f_2(\beta,E_{\mathbf{k}},\zeta)]  \label{eq:q} \\
 R &= \frac{1}{2 \vert \Psi \vert^2} \int \frac{d\mathbf{k}}{(2 \pi)^3} \left[ \frac{ f_1(\beta,E_{\mathbf{k}},\zeta) + (E_{\mathbf{k}}^2 - 3 \xi_{\mathbf{k}}^2) f_2(\beta,E_{\mathbf{k}},\zeta) }{3 \vert \Psi \vert^2} \right. \notag \\*
 & \hspace{7em} + \left. \frac{4(\xi_{\mathbf{k}}^2 - 2 E_{\mathbf{k}}^2)}{3} f_3(\beta,E_{\mathbf{k}},\zeta) + 2 E_{\mathbf{k}}^2 \vert \Psi \vert^2 f_4(\beta,E_{\mathbf{k}},\zeta) \right]  \label{eq:r}
 \end{align}
 The functions $f_j(\beta,\epsilon,\zeta)$ in the above expressions are defined by
 \begin{align}
 f_{j}(\beta,\epsilon,\zeta)=\frac{1}{\beta}\sum_{n}\frac{1}{\left[\left(\omega_n-i\zeta\right)^2+\epsilon^2\right]^j}
 \end{align}
 with the fermionic Matsubara frequencies $\omega_n=(2n+1)\pi/\beta$.
 In this treatment, the chemical potentials of the two pseudo-spin species $\mu_{\uparrow}$ and $\mu_{\downarrow}$ are combined into the average chemical potential $\mu = (\mu_{\uparrow} + \mu_{\downarrow})/2$ and the  imbalance chemical potential $\zeta = (\mu_{\uparrow} - \mu_{\downarrow})/2$, the latter determining the difference between the number of particles in each spin-population. The quantity $\xi_{\mathbf{k}} = \frac{k^2}{2m} - \mu$ is the dispersion relation for a free fermion, $E_{\mathbf{k}} = (\xi_{\mathbf{k}}^2 + \vert \Psi_{\mathbf{x},\tau} \vert^2)^{1/2}$ is the local Bogoliubov excitation energy, and $a_s$ is the $s$-wave scattering length that determines the strength and the sign of the contact interaction. In absence of spatial and temporal variations, the thermodynamic potential $\Omega_s$ determines the value of the pair-breaking gap $\vert \Psi_{\infty} \vert$ for the uniform system through the saddle-point gap equation
 \begin{equation}
 \frac{\partial \Omega_s}{\partial \vert \Psi \vert^2} \Psi = 0
 \label{eq:gap}
 \end{equation}
 This equation is solved self-consistently together with the number equation to obtain the correct values of $\vert \Psi_{\infty} \vert$ and $\mu$ for a given set of system parameters.

 In principle, all expansion coefficients \eqref{eq:omega}--\eqref{eq:r} fully depend upon the order parameter $\Psi(\mathbf{x},\tau)$, but in practice, we assume that the coefficients associated with the second order derivatives of the pair field can be kept constant and equal to their bulk value, since retaining their full space-time dependence would lead us beyond the second-order approximation of the gradient expansion. This means that in expressions \eqref{eq:c}, \eqref{eq:e}, \eqref{eq:q} and \eqref{eq:r} for the coefficients $C$, $E$, $Q$ and $R$, we set $\vert \Psi(\mathbf{x},\tau) \vert^2 \rightarrow \vert \Psi_{\infty} \vert^2$ and $E_{\mathbf{k}} \rightarrow E_{\mathbf{k},\infty} =  (\xi_{\mathbf{k}}^2 + \vert \Psi_{\infty} \vert^2)^{1/2}$. For the thermodynamic potential $\Omega_s$ and the coefficient $D$, on the other hand, the full space-time dependence of the order parameter is preserved.

 From the Euclidian-time action functional \eqref{eq:action}, the EFT equation of motion for the pair field $\Psi(\mathbf{r},t)$ is found to be
 \begin{equation}
 i \tilde{D}(\vert \Psi \vert^2) \frac{\partial \Psi}{\partial t} = -C \, \nabla_{\mathbf{r}}^2 \Psi + Q \frac{\partial^2 \Psi}{\partial t^2} + \left( \mathcal{A}(\vert \Psi \vert^2) + 2  E \, \nabla_{\mathbf{r}}^2 \vert \Psi \vert^2 - 2  R \frac{\partial^2 \vert \Psi \vert^2}{\partial t^2} \right) \Psi \label{eq:Appeqofmot}
 \end{equation}
 where the coefficients $\tilde{D}$ and $\mathcal{A}$ are defined as
 \begin{align}
  \tilde{D}
 =\frac{\partial\left(  |\Psi|^2 D  \right)  }{\partial
 \left(|\Psi|^2\right)}  \qquad \mathcal{A}  &  =\frac{\partial\Omega_{s} }{\partial \left( |\Psi|^2 \right)} \label{eq:A&D}
 \end{align}
 The first term on the right-hand side of the equation can be identified as a kinetic energy term, while the non-linear term represents a system-inherent potential for the field. The ratio $\tilde{D}/C$ can be interpreted as a renormalization factor for the mass of the fermion pairs \cite{THKTVPrA94} and the coefficient $\mathcal{A}$ determines the uniform background value of the system, since $\mathcal{A}(\Psi) \, \Psi = 0$ is nothing but the aforementioned gap equation \eqref{eq:gap}. It has been verified that in the deep BEC-limit $\left( 1/k_F a_S \gg 1 \right)$, the equation correctly tends to the Gross-Pitaevskii equation for bosons with a mass $M=2m$ and an s-wave boson-boson scattering length $a_B = 2 \, a_s$ \citep{THLombardiPhD}.

 \subsection*{Variational derivation of the healing length}
 \label{sec:appxi}

 \begin{figure}[htbp]
 \centering
 \centerline{
 \includegraphics[scale=0.8]{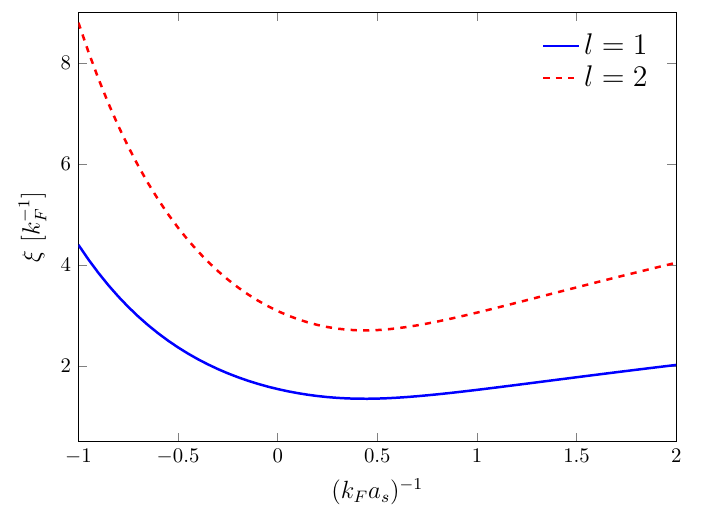}}
 \caption{Variational estimate of the vortex width $\xi$ in function of the interaction parameter, for winding numbers $l=1$ and $l=2$.}
 \label{fig:xivar}
 \end{figure}

 We can derive an analytic expression for the healing length $\xi$ associated to the width of a stationary vortex in a Fermi superfluid by considering a variational ansatz for the wavefunction and minimizing the free energy of the system. A popular model to describe the pair field of the stationary vortex state is the hyperbolic tangent function:
 \begin{equation}
 \label{eq:TanhVor}
 \Psi(r,\phi) = \vert \Psi_{\infty} \vert \tanh\left(\frac{r}{\sqrt{2} \xi} \right) \, e^{i l \phi}.
 \end{equation}
 The EFT free energy functional $F$ in terms of the polar coordinates $r$ and $\phi$ is given by
 \begin{multline}
 F[\Psi] = \int d\phi \int_0^{\infty} r \, \mathrm{d}r  \left[ X(\vert \Psi \vert^2) + \tilde{C} \, \left( \partial_r \bar{\Psi} \, \partial_r \Psi + \frac{1}{r^2} \partial_{\phi} \bar{\Psi}\,  \partial_{\phi} \Psi \right) - \frac{\tilde{E}}{2} \left( (\partial_r \vert \Psi \vert^2 )^2 + \frac{1}{r^2} (\partial_{\phi} \vert \Psi \vert^2 )^2 \right) \right],
 \label{eq:Fpolar}
 \end{multline}
 with
 \begin{equation}
 X(\vert \Psi \vert^2) = \Omega_{s}(\vert \Psi \vert^2) - \Omega_{s}(\vert \Psi_{\infty} \vert^2).
 \end{equation}
 The subtraction of the term $\Omega_{s}(\vert \Psi_{\infty} \vert^2)$ indicates that the energy is measured with respect to the energy of the uniform system.  By substituting the ansatz \eqref{eq:TanhVor} for the pair field into the free energy and making a change of integration variable $u = x / (\sqrt{2} \xi)$, we obtain
 \begin{multline}
 F = 4 \pi \int_{0}^{\infty} u \, \mathrm{d}u \left[ \xi^2 \, X(u)  + \frac{\tilde{C} \, \vert \Psi_{\infty} \vert^2}{2} \text{sech}^4\left( u \right) + \frac{\tilde{C} \, \vert \Psi_{\infty} \vert^2 \, l^2}{2 \, u^2} \tanh^2 \left( u \right)  - \tilde{E} \, \vert \Psi_{\infty} \vert^4 \, \text{sech}^4\left(u \right) \, \tanh^2 \left( u \right) \right],
 \label{eq:Ftanhu}
 \end{multline}
 The integral over the term with $X(u)$ converges, but has to be calculated numerically. The second and fourth integral also converge, and can be calculated exactly. The integral of the third term, on the other hand, yields a logarithmic divergence. However, the main quantity of interest for the variational treatment is the derivative of the free energy with respect to $\xi$, which, in contrast to the free energy itself, does not diverge. One then obtains \cite{THVerhelstPhysC}
 \begin{align}
 \frac{d F}{d \xi} &= 8 \, \pi \, \xi  \int_{0}^{\infty} u  \, X(u) \, \mathrm{d}u  -  \frac{4  \pi \, \tilde{C} \, \vert \Psi_{\infty} \vert^2 \, l^2}{\xi} \int_{0}^{\infty} \tanh \left( u \right) \, \text{sech}^2 \left( u \right) \mathrm{d}u \\
 &= 8 \, \pi \, \xi_v  \int_{0}^{\infty} u  \, X(u) \, \mathrm{d}u  -  \frac{2  \pi \, \tilde{C} \, \vert \Psi_{\infty} \vert^2 \, l^2}{\xi}.
 \end{align}
 By setting the above equation equal to zero, we find the following variational expression for the vortex width:
 \begin{equation}
 \label{eq:XiVorVar}
 \xi = \frac{1}{2} \sqrt{\frac{\tilde{C} \, \vert \Psi_{\infty} \vert^2 \, l^2}{B}},
 \end{equation}
 with
 \begin{equation}
 B = \int_{-\infty}^{\infty} X(u) \, u \, du.
 \end{equation}
 Figure \ref{fig:xivar} shows the behavior of this quantity in function of the interaction parameter $(k_F a_s)^{-1}$ for $l=1$ and $l=2$. A more extensive study on the healing length of a fermionic superfluid across the BEC-BCS crossover can be found in Ref.\ \cite{THPalestiniStrinati}.

 \subsection*{Linearization of the equation of motion}
 \label{sec:applin}

 To describe small fluctuations of the pair field, we add a  perturbation field $\Phi(\mathbf{r},t)$ to the stable vortex solution $\Psi_s(\mathbf{r})$:
 \begin{equation}
 \Psi(r,\phi,z,t) = \Big( f(r) + \Phi(r,\phi,z,t) \Big) e^{i l \phi}.
 \end{equation}
 This perturbed solution can be substituted into the EFT equation of motion \eqref{eq:Appeqofmot}, which can then be linearized with respect to the perturbation field. This requires the coefficients $\tilde{D}$ and $\mathcal{A}$ (which depend on the local value of the order parameter) to be expanded around the stationary solution:
 \begin{align}
 \tilde{D}(\vert \Psi \vert^2) &= \tilde{D}_s + f(r) \left[\Phi(x,y,t) +  \Phi^*(x,y,t) \right] \partial_s\tilde{D}_s  + \cdots, \\[5pt]
 \mathcal{A}(\vert \Psi \vert^2) &= \mathcal{A}_s + f(r) \left[ \Phi(x,y,t) + \Phi^*(x,y,t) \right] \partial_s \mathcal{A}_s  + \cdots.
 \end{align}
 Here, we have used the notations
 \begin{equation}
 f_s = f \big(\vert \Psi_s(x) \vert^2 \big), \qquad \partial_s f_s = \frac{\partial f}{\partial \vert \Psi \vert^2} \bigg\vert_{\vert \Psi_s \vert^2} .
 \end{equation}
 Small excitations of the system can be described by assuming a plane-wave solution for the fluctuation field of the form \cite{THLund2}
 \begin{equation}
 \Phi(r,\phi,z,t) = \phi_1(r) e^{i(m\phi + k_z z - \omega t)} + \phi_2^*(r) e^{-i(m\phi + k_z z - \omega^* t)},
 \end{equation}
 where $m$ is an angular momentum quantum number (relative to the quantum number $l$ of the condensate) and $k_z$ is the wave number along the symmetry axis of the stationary vortex solution.
 After substituting this ansatz into the equation of motion, terms of equal order in the perturbation amplitudes can be collected. The first order terms result in two coupled linear differential equations for the perturbation amplitudes $\phi_1(r)$ and $\phi_2(r)$:
 \begin{align}
 \label{eq:AppDQuv1}
 \alpha_1(r) \frac{\partial^2 \phi_1}{\partial r^2} + &\alpha_2(r) \frac{\partial \phi_1}{\partial r} + \Big( \omega^2 \, \alpha_3(r) + \omega \, \alpha_4(r) + \alpha_{5,+}(r) \Big) \phi_1 \notag \\
 &+ \alpha_6(r) \frac{\partial^2 \phi_2}{\partial x^2} + \alpha_7(r) \frac{\partial \phi_2}{\partial r} + \Big( \omega^2 \, \alpha_8(r) + \alpha_9(r) \Big) \phi_2 = 0,
 \end{align}
 \begin{align}
 \label{eq:AppDQuv2}
 \alpha_1(r) \frac{\partial^2 \phi_2}{\partial r^2} + &\alpha_2(r) \frac{\partial \phi_2}{\partial r} + \Big( \omega^2 \, \alpha_3(r) - \omega \, \alpha_4(r) + \alpha_{5,-}(r) \Big) \phi_2 \notag \\*
 &+ \alpha_6(r) \frac{\partial^2 \phi_1}{\partial r^2} + \alpha_7(r) \frac{\partial \phi_1}{\partial r} + \Big( \omega^2 \, \alpha_8(r) + \alpha_9(r) \Big) \phi_1 = 0,
 \end{align}
 where the coefficients $\alpha_j(r)$ are given by
 \begin{align}
 \alpha_1(r)  &= \tilde{C} - \tilde{E} \, f^2(r), \\
 \label{eq:alpha1vor}
 \alpha_2(r) &= \frac{\tilde{C}}{r} -   \tilde{E} \, f(r) \frac{f(r) + 2 \, r \, f'(r)}{r}, \\
 \alpha_3(r) &= Q - \tilde{R} \, f(r)^2, \\
 \alpha_4(r) &= \tilde{D}_s, \\
 \alpha_{5,\pm}(r) &= -\frac{\tilde{C} (l \pm m)^2}{r^2} + \tilde{E} \, f(r) \frac{m^2 \, f(r) - 3 \, r \, f'(r)}{r^2}  - \Big( \tilde{C} - \tilde{E} f^2(r) \Big) \, k_z^2, \notag \\
 & \quad - \mathcal{A}_s(r)  - \partial_s \mathcal{A}_s(r) \, f^2(r) + \tilde{E}  \Big( 2(f'(r))^2 - 3 \, f(r) \, f''(r) \Big), \\
 \alpha_6(r) &= -\tilde{E} \, f^2(r), \\
 \alpha_7(r) &=  -\tilde{E} \, f(r) \frac{f(r) + 2 \, r \, f'(r)}{r}, \\
 \alpha_8(r) &= -\tilde{R} \, f^2(r), \\
 \alpha_9(r) &= \tilde{E} \, f^2(r) \, k_z^2 - \partial_s \mathcal{A}_s(r) \, f^2(r) + \tilde{E} \, f(r) \frac{m^2 \, f(r) - r \, f'(r)}{r^2} \notag \\
 & \quad - \tilde{E} \, f(r) \, f''(r).
 \label{eq:alpha9vor}
 \end{align}

 \subsection*{Discretization and evolution of the equation of motion}
 \label{sec:appdis}
 In this section we elaborate on how the EFT equation of motion \eqref{eq:Appeqofmot} is discretized and solved numerically using the explicit RK4 algorithm.
 We introduce a field $\phi(\mathbf{r},t)$ such that
 \begin{equation}
 \phi = \frac{\partial \Psi}{\partial t}
 \label{eq:dt1}
 \end{equation}
 and
 \begin{equation}
 \bar{\phi} = \overline{\frac{\partial \Psi}{\partial t}} = \frac{\partial \bar{\Psi}}{\partial t}
 \end{equation}
 Substituting this into the equation of motion and making use of the fact that
 \begin{equation}
 \frac{\partial^2 \vert \Psi \vert^2}{\partial t^2} = \bar{\Psi} \frac{\partial^2 \Psi}{\partial t^2} + 2 \frac{\partial \bar{\Psi}}{\partial t} \frac{\partial \Psi}{\partial t} + \Psi \frac{\partial^2 \bar{\Psi}}{\partial t^2}
 \end{equation}
 we have
 \begin{equation}
 i \tilde{D}(\vert \Psi \vert^2) \phi = -\tilde{C} \, \nabla_{\mathbf{r}}^2 \Psi + Q \frac{\partial \phi}{\partial t} + \left( \mathcal{A}(\vert \Psi \vert^2) + \tilde{E} \, \nabla_{\mathbf{r}}^2 \vert \Psi \vert^2 - \tilde{R} \left( \bar{\Psi} \frac{\partial \phi}{\partial t} + 2 \vert \phi \vert^2 + \Psi \frac{\partial \bar{\phi}}{\partial t} \right) \right) \Psi
 \label{eq:dtphidtphib}
 \end{equation}
 In order to get an equation of the form $\partial_t \phi = ... \,$, we take the complex conjugate of \eqref{eq:dtphidtphib}, find an expression for $\partial_t \bar{\phi}$ in function of $\partial_t \phi$ and substitute this back into \eqref{eq:dtphidtphib}, yielding
 \begin{align}
 \frac{\partial \Phi}{\partial t} &= \frac{1}{Q \, (Q - 2 \tilde{R} \vert \Psi \vert^2)}  \left[ -Q \left( \mathcal{A} + \tilde{E} \, \nabla_{\mathbf{r}}^2 \vert \Psi \vert^2 - 2 \tilde{R} \vert \phi \vert^2  \right) \Psi + i \tilde{D} \left( Q \phi - \tilde{R} \, \Psi \left( \bar{\phi} \Psi + \phi \bar{\Psi} \right) \right) \right. \notag \\
  &+ \left. \tilde{C} \left( \Psi^2 \tilde{R} \, \nabla_{\mathbf{r}}^2 \bar{\Psi}   + \nabla_{\mathbf{r}}^2 \Psi (Q - \tilde{R} \vert \Psi \vert^2)  \right) \right]
  \label{eq:dt2}
 \end{align}
 Equations \eqref{eq:dt1} and \eqref{eq:dt2} form a system of two coupled partial differential equations of the form:
 \begin{align}
 \frac{\partial \Psi}{\partial t} &= f(\phi) \\
 \frac{\partial \phi}{\partial t} &= g(\Psi,\phi)
 \end{align}
 where $f(\phi) = \phi$ and $g(\Psi,\phi)$ is given by \eqref{eq:dt2}. In the case of a 2D system, we use finite mesh widths $\Delta x$ and $\Delta y$ and a finite time step $\Delta t$ to discretize space-time into a grid of $L \times M \times N$ points by writing $x_l = l \Delta x$ with $l = 1,...,L$, $y_m = m \Delta y$ with $m = 1,...,M$ and $t_n = n \Delta t $ with $n = 1,...,N$. This allows us to approximate the spatial derivatives by central finite difference formulas:
 \begin{align}
 \frac{\partial^2 \Psi(x,y,t)}{\partial x^2} = \frac{\Psi_{l+1,m,n} - 2 \, \Psi_{l,m,n} + \Psi_{l-1,m,n}}{\Delta x^2} \\[10pt]
 \frac{\partial^2 \Psi(x,y,t)}{\partial y^2} = \frac{\Psi_{l,m+1,n} - 2 \, \Psi_{l,m,n} + \Psi_{l,m-1,n}}{\Delta y^2}
 \end{align}
 where we use the notation $\Psi_{l,m,n} = \Psi(x_l,y_m,t_n)$.
 \color{black}
 A hard wall potential, equal to $0$ for $r=\sqrt{x^2+y^2}<R$ and $V_0=10^4 \times E_F$ for $r\geq R$,
 was implemented for the field $\Psi$.
 \color{black}
 If we now know the values $\Psi_{l,m,n}$ and $\phi_{l,m,n}$ at a certain time step $t_n$ for all positions $x_l$ and $y_m$, the explicit RK4 method allows us to calculate for every position the values $\Psi_{l,m,n+1}$ and $\phi_{l,m,n+1}$ of the next time step by using the following algorithm  \citep{THSuliMayers}:
 \begin{align}
 &p_{1_{l,m,n}} = f(\phi_{l,m,n}) \\
 &p_{2_{l,m,n}} = g(\Psi_{l,m,n},\phi_{l,m,n}) \\
 &q_{1_{l,m,n}} = f(\phi_{l,m,n}+p_{2_{l,m,n}}/2) \\
 &q_{2_{l,m,n}} = g(\Psi_{l,m,n}+p_{1_{l,m,n}}/2,\phi_{l,m,n}+p_{2_{l,m,n}}/2) \\
 &r_{1_{l,m,n}} = f(\phi_{l,m,n}+q_{2_{l,m,n}}/2) \\
 &r_{2_{l,m,n}} = g(\Psi_{l,m,n}+q_{1_{l,m,n}}/2,\phi_{l,m,n}+q_{2_{l,m,n}}/2) \\
 &s_{1_{l,m,n}} = f(\phi_{l,m,n}+r_{2_{l,m,n}}) \\
 &s_{2_{l,m,n}} = g(\Psi_{l,m,n}+r_{1_{l,m,n}},\phi_{l,m,n}+r_{2_{l,m,n}}) \\
 &\Psi_{l,m,n+1} = \Psi_{l,m,n} + \frac{\Delta t}{6}(p_{1_{l,m,n}} + 2 \, q_{1_{l,m,n}} + 2 \, r_{1_{l,m,n}} + s_{1_{l,m,n}}) \\
 &\phi_{l,m,n+1} = \phi_{l,m,n} + \frac{\Delta t}{6}(p_{2_{l,m,n}} + 2 \, q_{2_{l,m,n}} + 2 \, r_{2_{l,m,n}} + s_{2_{l,m,n}})
 \end{align}
 This scheme can be repeated until the solution has been evolved up to the desired point in time.

\color{black}
\subsection*{Three dimensional effect}
\color{black}
We briefly consider the splitting instability in three dimensions. It has been demonstrated for the case of 3D BECs that, in the early stages of the decay, there might arise a periodic structure of alternating split and non-split regions along the $z$-direction of the vortex line, a so-called ``{chain-structure}'' \cite{THHuh2,THIsoshima}. This uneven splitting can make it difficult to compare experimental observations and theoretical predictions. In the context of the linear stability analysis, the presence of a third dimension is described by the wave number $k_z$, as introduced in expression \eqref{eq:planewave}. We have observed in our calculations that there is a critical value $k_{z,c}$ for this wave number above which no more unstable modes exist. Since the chain structure can only be induced if an unstable mode with a finite value of $k_z$ fits into the system, no three-dimensional deformations will occur if the system size along the vortex axis is smaller than $2 \pi/k_{z,c}$. In order to observe our predictions for the splitting instability \color{black} with $k_z=0$, \color{black} the thickness of the atomic clouds along the vortex line must be smaller than $2 \pi/k_{z,c} \approx 20\xi$, $8\xi$, $4\xi$ for $(k_F a_s)^{-1} = -2,0,2$ respectively, according to our numerical analysis. In the BEC- and crossover regime, the value of $k_{z,c}$ will slightly depend on the radial system size $R$.


\end{document}